\documentclass[%
 aip,
 pop
 amsmath,amssymb,
 reprint,%
]{revtex4-1}

\usepackage{graphicx}
\usepackage{dcolumn}
\usepackage{bm}

\newcommand{\sss}{\scriptscriptstyle}
\newcommand {\be}{\begin{equation}} 
\newcommand{\ee}{\end{equation}}    

\def\vti{v_{{\sss T}i}}

\def\vte2{v_{{\sss T}\alpha}}

\begin{document}
\preprint{AIP/123-QED}

\title[ ]{ Alfv\'{e}n wave coupled with flow-driven fluid instability in interpenetrating plasmas }

\author{J. Vranjes }
  \affiliation{
Instituto de Astrofisica de Canarias, 38205 La Laguna, Tenerife, Spain and\\
Departamento de Astrofisica, Universidad de La Laguna, 38205 La Laguna, Tenerife, Spain.
}%

\date{\today}

\begin{abstract}

 The Alfv\'{e}n wave is analyzed  in case of one quasineutral plasma propagating  with some constant  speed $v_0$ through another static quasineutral plasma. A dispersion equation is derived describing the Alfv\'{e}n wave coupled with the flow driven mode $\omega= k v_0$  and solutions are  discussed analytically and numerically. The usual solutions for two oppositely propagating  Alfv\'{e}n waves are substantially modified due to the flowing plasma. More profound is modification of the solution propagating in the negative direction with respect to the magnetic field and the plasma flow. For a large enough flow speed (exceeding the Alfv\'{e}n speed in the static plasma), this negative  solution  may become non-propagating, with frequency equal to zero. In this case it represents a spatial variation of the electromagnetic field. For greater flow speed it becomes a forward mode, and it may  merge with the positive one. This merging of the two modes  represents the starting point for a flow-driven  instability, with two complex-conjugate solutions. The Alfv\'{e}n wave in interpenetrating plasmas is thus modified and coupled with the flow-driven mode and this coupled mode is shown to be  growing when the flow speed is large enough. The energy for the instability is macroscopic kinetic energy of the flowing plasma. The dynamics of plasma particles caused by such a coupled  wave still remains similar to the ordinary Alfv\'{e}n wave. This means that well-known stochastic heating by the Alfv\'{e}n wave may work, and this  should additionally support the potential role of the   Alfv\'{e}n wave in the coronal heating.
\end{abstract}

\pacs{52.35.Bj, 52.30.-q, 96.60.-j}
\maketitle

\section{Introduction}

Alfv\'{e}n wave (AW) has been a hot topic in plasma physics ever since its discovery  \citep{alf}, both in the general plasma
theory and laboratory plasmas \citep{al,tan, bar, hol3, ros, wat, gig, bas, v4, v5}, and in astrophysics \citep{kp, pud, gaz}.
This is even more so  for solar atmosphere which is a perfect natural laboratory for various plasma waves and instabilities, and almost any of the
plasma modes observed in the laboratory or predicted by theory may be expected and/or observed there. 

The Alfv\'{e}n
wave is widely believed to play an important role in the heating of upper layers in the solar atmosphere. Indeed, the wave  has apparently been detected in numerous observations \citep{jes, dp2}, and it has been studied in a great amount of works in the past (to mention  just a few like for example
 \citep{hol, hol2, lu,   fl,  mur}). In fact this is by far the most studied mode in the solar plasma environment.

In view of its potential role in the coronal heating, and this even at frequencies far below the particle gyrofrequency, known as non-resonant (stochastic) heating \citep{wang, lu, wu}, it is of great importance to search for reliable sources for the the AW and to be able to distinguish it from  some other electromagnetic (EM) perturbations and instabilities which may exhibit similar behavior. 

One of such EM instabilities develops when one flowing (f) plasma with its specific set of parameters propagates through another static (s) plasma whose parameters may in principle be different. Both plasmas separately may be  quasi-neutral, or the quasi-neutrality applies for the system as a whole (both situations have been studied in the literature)  and they may have different densities $n_f$ and $n_s$. The indices $f, s$ will be used to denote parameters in the two respective plasmas. 

Within kinetic theory,  in our recent papers  \citep{v1, v2, v3, v4},  we have shown that a new kind of instability may develop in such systems. The instability discussed for longitudinal electrostatic perturbations is purely kinetic by nature and  current-less (the latter meaning that the plasma contains no currents in the initial state). For such modes, the real part of the wave frequency is mainly determined by the static plasma parameters, while  the instability is mainly due to ions from the flowing plasma, and these features make it very different from classic current-carrying plasmas where the instability is caused by electrons passing through static ions. 

When one  population of particles moves as a whole through a plasma, there may be a collective mode associated with this motion \citep{ichi}, and an instability may develop \citep{dok} even in fluid theory, and  both  real and imaginary parts of the wave frequency may be completely determined by the flow. The properties of such flow-driven (FD) instabilities and  modes may sometimes be similar to normal plasma waves,  and some care is needed to make a clear distinction between them. 

In the present work we shall describe some features of an electromagnetic FD instability coupled with the AW within fluid theory, and consequences of this coupling on the  AW.  In such a system of two mutually coupled plasmas (one flowing and another static) the usual AW cannot be expected in any of the two plasmas separately. Instead, it will be shown that  a coupled mode develops with some properties (e.g., dynamics of particles involved in the wave) similar to the usual AW, but the usual AW  may  become modified by the flow even for relatively small flow speed. For a fast enough flow the resulting mode may be unstable and  completely  determined by the flow, and yet (for some specific cases of the plasma flow) the wave frequency may still be exactly equal to the usual AW frequency of the static plasma (see later in the text and Figs.~\ref{f1},~\ref{f2}), although in most cases it is different. 
 
 Unaware of the presence of the flowing plasma, and using  the  usual AW as a tool  for remote diagnostics,  observers  might  come out with rather inaccurate estimates regarding the values of local plasma parameters. The purpose of this work is to point out what may be expected in such coupled plasmas and how to interpret observations appropriately.

\section{Basic equations and model}

\begin{figure}
   \centering
  \includegraphics[height=6.5cm,bb=14 51 136 160,clip=]{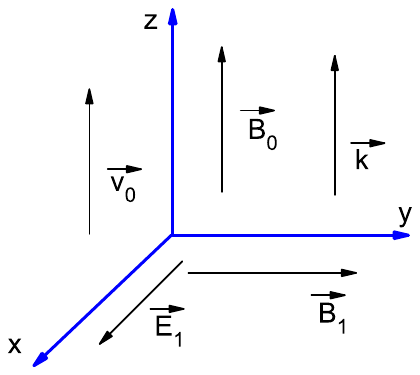}
      \caption{Geometry of the EM wave propagating in the direction of the magnetic field.  \label{m}}
       \end{figure}

We  shall assume one flowing (f) electron-ion plasma moving with the speed $\vec v_0= v_0\vec e_z$ through another static (s) electron-ion plasma plasma, and we shall use standard equations appropriate for electromagnetic perturbations. Each of the  plasmas is  initially separately currentless. The flow is  in the direction of a background magnetic field $\vec B_0=B_0\vec e_z$, see Fig.~\ref{m}.

The perturbations are in the low-frequency limit $\omega\ll \Omega_i=e B_0/m_i$, the ion mass in both plasmas is the same and for the two systems we have separate quasineutrality conditions satisfied  $n_{fi0}=n_{fe0}=n_{f0}$, $n_{si0}=n_{se0}=n_{s0}$. In such a regime the perturbations may be assumed as linearly polarized \citep{ch} and we may assume that the perturbed magnetic field is $\vec B_1=B_y\vec e_y$, similar to the case of an ordinary AW. Perturbations are assumed of the shape $\sim\exp(-i \omega t+ i k z)$ and we use the usual MHD equations for the two plasmas.  From the Faraday law we  have
\[
\vec E_1=\frac{\omega B_y}{k} \vec e_x.
\]
We further use Ohm's law and omit  all unessential terms like electron mass corrections, collisions, and Hall effect (details about the latter can be found elesewhere\citep{pan}), and for the static plasma
we have $\vec E_1+ \vec u_1\times \vec B_0=0$, which yields
\be
\vec u_1=-\frac{\vec B_0\times \vec E_1}{B_0^2}=-\frac{\omega}{k}\frac{B_y}{B_0}\vec e_y.\label{e1}
\ee
For the flowing plasma we have $\vec E_1+ \vec v_1\times \vec B_0+ \vec v_0\times \vec B_1=0$, which yields
\be
\vec v_1=-\frac{\vec B_0\times \vec E_1}{B_0^2} + \frac{v_0}{B_0} \vec B_1=-\frac{\omega}{k}\frac{B_y}{B_0}\vec e_y + v_0\frac{B_y}{B_0}\vec e_y.\label{e2}
\ee
 The Amp\`{e}re law for the present case of the two plasmas reads:
 \be
 \nabla\times \vec B=\mu_0(\vec j_s+ \vec j_f). \label{e3}
 \ee
The currents $\vec j_{s, f}$ can be calculated from the MHD momentum equations
\be
\rho_s\frac{\partial\vec u_1}{\partial t}=\vec j_{s1}\times \vec B_0, \label{e4}
\ee
\be
\rho_f\left(\frac{\partial}{\partial t} + \vec v_0\cdot\nabla\right)\vec v_1=\vec j_{f1}\times \vec B_0, \label{e5}
\ee
where $\rho_j=(m_e+ m_i)n_j$. This yields the dispersion equation for low frequency EM perturbations propagating in the direction of the magnetic field
\be
\omega^2\left(1+ \frac{v_{{\sss A}f}^2}{v_{{\sss A}s}^2}\right) - 2 k v_0 \omega + k^2 \left(v_0^2- v_{{\sss A}f}^2\right)=0.
\label{e6}
\ee
Here  $v_{{\sss A}j}^2=B_0^2/(\mu_0 \rho_j)$ are the Alfv\'{e}n speeds for the two separate plasmas. Eq.~(\ref{e6}) describes coupled AW and flow driven (FD) mode. The  frequency of the propagating perturbations is
\be
\frac{\omega}{k}=\frac{1}{1+ n_s/n_f} \left\{v_0\pm v_{{\sss A}s}\left[\frac{n_s}{n_f}\left(1+ \frac{n_s}{n_f} -\frac{v_0^2}{v_{{\sss A}s}^2}\right)\right]^{1/2}\right\}. \label{e7}
\ee
Further in the text, the two solutions with plus and minus  signs will be called P and N solutions, respectively. Quite generally,  there can be no instability as long as $v_0\leq v_{{\sss A}s}$. However, there may be an instability  on condition
\be
v_0^2>v_{{\sss A}s}^2 \left(1+ \frac{n_s}{n_f}\right)= v_{{\sss A}s}^2+ v_{{\sss A}f}^2. \label{e8}
\ee
From  Eq.~(\ref{e7}) it can also be concluded that the instability is absent if the density of the flowing plasma is small, i.e., if the following condition is satisfied:
\be
\frac{n_f}{n_s}\leq \frac{1}{v_0^2/v_{{\sss A}s}^2 -1}. \label{e9}
\ee
  The features of the  growing solutions are  discussed further in the text.

 \section{Properties of solutions in the presence of  flowing plasma}

\begin{figure}
   \centering
\includegraphics[height=6.5cm,bb=22 16 281 214,clip=]{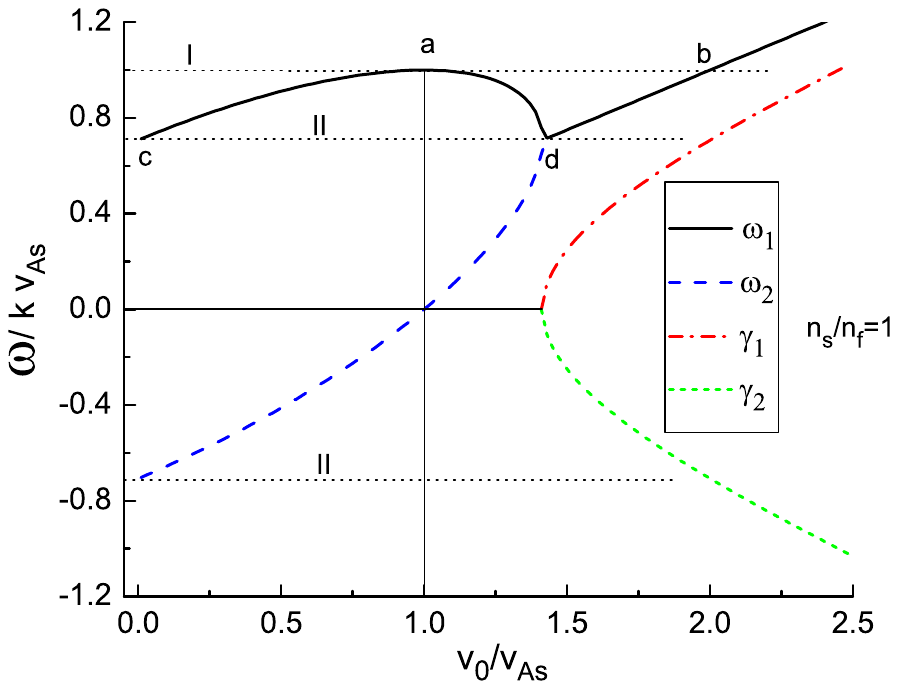}
\caption{Two solutions given by  Eq.~(\ref{e7}) for  two interpenetrating plasmas of the same density: normalized frequency in terms of the normalized speed of the flowing plasma. \label{f1}}
\end{figure}

\begin{figure}
   \centering
\includegraphics[height=6.5cm,bb=22 16 284 215,clip=]{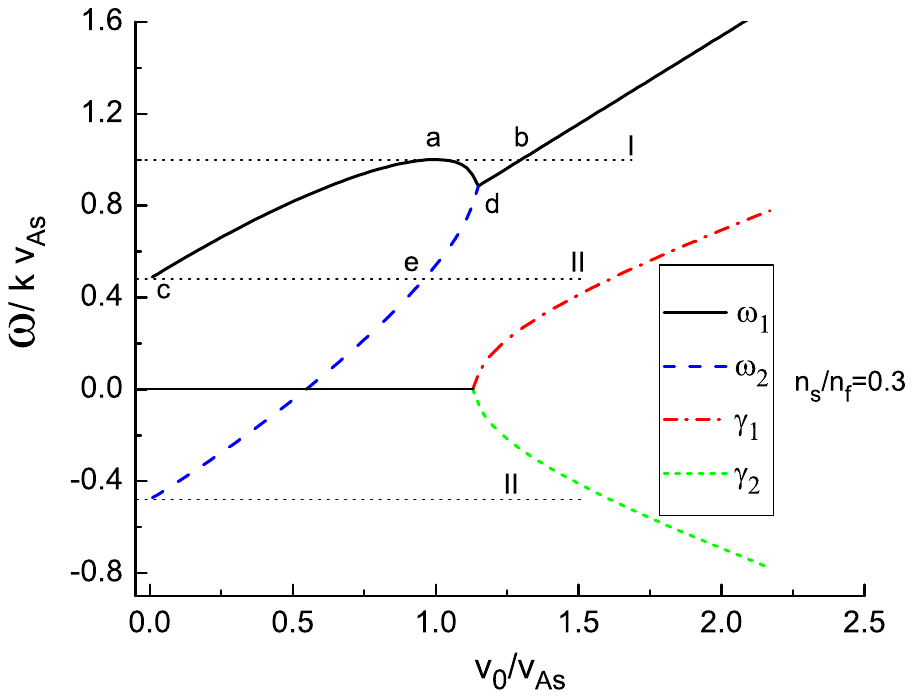}
\caption{The same as in Fig.~\ref{f1} but for $n_s/n_f=0.3$. \label{f2}}
\end{figure}

Without the flowing plasma, from Eq.~(\ref{e7}) we have  the AW in the static plasma $\omega_s=k v_{{\sss A}s}$. In the presence of both plasmas but with a negligible flow we have $\omega_{s+f}= B_0/[\mu_0 m (n_s+ n_f)]^{1/2}$.  These two characteristic limits  are presented by dotted lines I,~II in Figs.~\ref{f1}, \ref{f2}. Only in these particular limits  one may speak about the ordinary AW,  the normalized frequency is constant, hence the horizontal lines I,~II.  In the  figures, for the densities $n_s/n_f= 1,\, 0.3$, respectively, the line II has values  $0.71,\, 0.48$, and this combined AW frequency  $\omega_{s+f}$ is marked by the letter $c$ in figures.

However, for any finite value of $v_0$, the wave frequency is modified by the flow. For the two density ratios $n_s/n_f$, the solution becomes complex, $\omega=\omega_r+ i \gamma$, for $v_0>v_{{\sss A}s}$ greater than  $1.41$ and  $1.15$ respectively, for parameters in Figs.~\ref{f1},~\ref{f2}, as can be concluded from Eq.~(\ref{e8}), i.e.,  the threshold is reduced for a more dense flowing plasma. This is logical as it implies more energy stored in the flow. It is interesting to observe  that in the range of the stable mode ($\gamma=0$) the real frequency for the P-solution  in combined two plasmas may become equal to the AW frequency in static plasma $\omega_{s}$; this is a direct consequence of the flow due to which the frequency is increased up to the line I. One of these  situations is marked by letter $a$ in Figs.~\ref{f1},~\ref{f2}.
   \begin{figure}
   \centering
\includegraphics[height=6.5cm,bb=22 16 278 220,clip=]{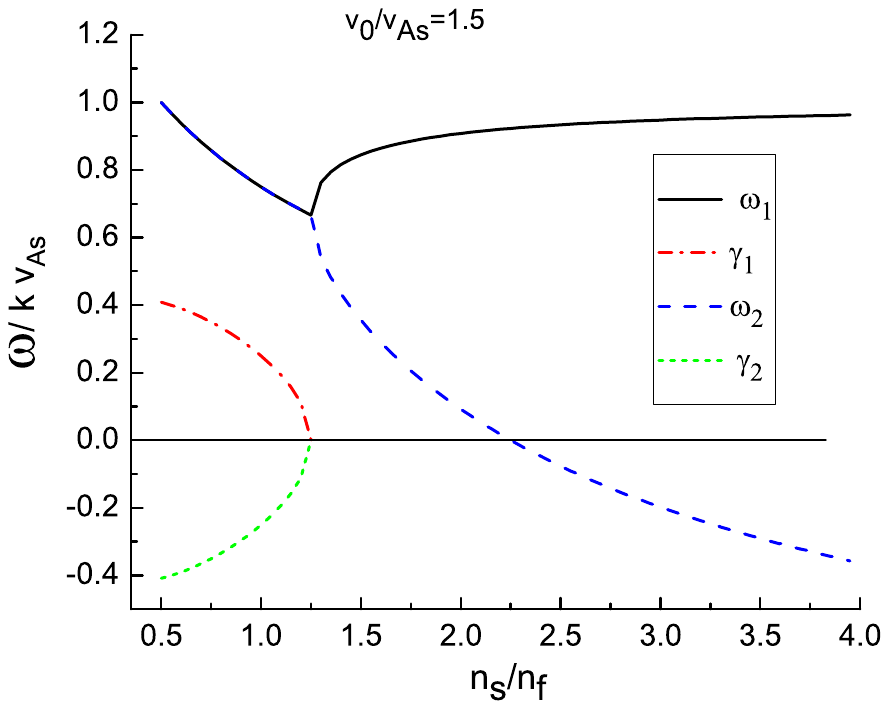}
\caption{Vanishing of instability for  reduced density of the flowing plasma, and transformation of the N solution from negative to forward, in terms of the density ratio of two plasmas. \label{f4}}
\end{figure}

Note  that in this same case the frequency for the N-solution for parameters in Fig.~\ref{f1} passes through zero. This implies that there is no propagation, and we have just a spatial variation of the magnetic field, with the spatial length $2 \pi/k$.  In case when the flowing plasma is heavier, this point of passing through zero frequency is shifted towards smaller values of $v_0$ as Fig.~\ref{f2} shows.  It is remarkable that in general, the N-solution becomes a forward mode and it can have practically any frequency in the range from $-\omega_{s+f}$ up to nearly $\omega_s$ as can be concluded from Figs.~\ref{f1},~\ref{f2}. Hence, if flow effects are not taken into account properly,  there exists an obvious possibility for misidentification  of modes and for mistakes in estimate of plasma parameters when wave characteristics are used for remote plasma probing. Further, the P- and N-solutions become coupled when the latter becomes a forward wave and when it is increased enough, and this is the reason for decreasing line of P-solution in the range $a-d$.

From quadratic dispersion equation (\ref{e6}) it is easily seen that  two identical real solutions exist (denoted by letter $d$ in Figs.~\ref{f1},~\ref{f2}) whenever
\be
v_0=\left(v_{{\sss A}s}^2 + v_{{\sss A}f}^2\right)^{1/2}=v_{{\sss A}s}(1+ n_s/n_f)^{1/2}.\label{e10}
\ee
This point represent the place where  P and N solutions merge. Also, whenever the densities of the two interpenetrating plasmas are equal, the frequency of the resulting flow-modified  mode at this merging point $d$ equals $\omega_{s+f}$. But this is only for $n_s/n_f=1$, observe that in Fig.~\ref{f2} the point  $d$ is  off the line II. The  point $e$ introduced in Fig.~\ref{f2} denotes  the place where N-solution  line (which is also a forward mode in this domain of $v_0$) crosses the line II.

Equally peculiar is the frequency marked by letter $b$, which is in the unstable mode regime when there is only one pair of complex-conjugate solutions. In that range both real and imaginary parts  of the frequency are determined by the flow, yet the real frequency part is equal to  $\omega_s$. This means that the solution  can easily be misidentified as the usual AW although clearly in this case one would have no explanation for the wave growth.
So a possibility for misidentification of modes exists  in both stable and unstable limits and for both P and N solutions; those are the stable point $a$ and unstable $b$ and they are relevant to AW $\omega_s$, and those are all points of the N-mode when its frequency is positive (including the stable point $d$).

The features related to condition (\ref{e9}) are presented in Fig.~\ref{f4}. For a fixed value $v_0$  the density ratio $n_s/n_f$ in the graph is allowed to  change and the instability vanishes for small enough flowing plasma density $n_f$; in the present case this is for $n_f < 0.8 n_s$. Thus the instability preferably develops when a  more dense plasma is accelerated into a dilute one. The N solution becomes forward $\omega_2>0$ at around $n_f=0.44 n_s$.

Finally, it should be stressed  that for any given wave-number $k$ we may have several possible flow speeds that give exactly the same real part of the wave-frequency. In Fig.~\ref{f1}, any real part of the P-frequency $\omega_1$ between lines I and II may be obtained for no less than three  different values of $v_0$ (with one of the three cases being unstable), while in Fig.~\ref{f2} this is narrowed to frequencies above those corresponding to the point $d$ and below line I. On the other hand, in Fig.~\ref{f4} it is seen that the same real frequency $\omega_1$ may be obtained (for any $k$)  for two different values of the ratio $n_s/n_f$. So there is an obvious ambiguity in determining plasma parameters  through the wave frequency.

\section{Discussions}

When one plasma enters another plasma and propagates through it, an electromagnetic Alfv\'{e}nic instability may develop if the speed
of the flowing plasma exceeds the Alfv\'{e}n speed of the static plasma, and this appears as a {\em necessary} instability condition. The supply energy for the instability is the kinetic energy of the streaming plasma. We have demonstrated that the instability may develop only when the density of the streaming plasma exceeds some critical value. The {\em sufficient} instability conditions are given by Eqs.~(\ref{e8}, \ref{e9}).

The AW behavior in the presence of a flowing (with speed $v_0$)  plasma can be summarized as follows: a) as long as $v_0^2/v_{{\sss A}s}^2<1+ n_s/n_f$ there are two modified solutions for the AW, one is always  positive (P) and the other is negative (N) for small $v_0$, b) due to the flow the N solution may propagate in the positive direction (note that we keep it calling N solution although it becomes positive, only in order to distinguish it from the P solution) and for $v_0^2/v_{{\sss A}s}^2=1+ n_s/n_f$ the two solutions merge, c) for $v_0^2/v_{{\sss A}s}^2>1+ n_s/n_f$ there are two complex-conjugate solutions.
 
 In general, more profound effects of the flow are on the N solution, as may be concluded from Figs.~\ref{f1},~\ref{f2}. The P solution phase speed increases till it reaches $v_{{\sss A}s}$,  but for a greater flow speed  it decreases (the region between the points $a$ and $d$ in Figs.~\ref{f1},~\ref{f2}). This happens in the range $1+ \epsilon<v_0^2/v_{{\sss A}s}^2<1+ n_s/n_f$  where $\epsilon$ can be any number  in the range  $0<\epsilon<n_s/n_f$.

Quite generally, the instability is expected to work well in an environment with relatively weak magnetic field. In fast expanding astrophysical clouds (which are typically more dense than the surrounding interstellar space) the instability should be ubiquitous.

Throughout the text, frequency and growth rate are calculated normalized to AW frequency in the static plasma $k v_{{\sss A}s}$ so the conclusions are general and valid for any wavelength. However, for  frequencies in the range of ion gyro-frequency or even higher, circular polarization should be taken into account. Such a generalization and coupling with ordinary and extraordinary EM waves can easily be done.  
  
  Various effects may to some extent modify (or reduce)  the  instability, like viscosity which can easily be included and it will consequently affect the wave mainly at shorter wavelengths \citep{v5}, and friction. In the corona itself the ion-ion collisions are rare \citep{vaa06} and the collision frequency is around 0.07 Hz. In the case of one ion species coming from lower layers and colliding with coronal ions,  the corresponding expressions for the collision frequency in such a situation are available in the literature\citep{vk08}, and the collision frequency is increased to a few Hz. However, the instability discussed here can clearly be very strong and these effects are not expected to play a major role.

Important  points to discuss are:  i) how realistic  the mode and instability discussed here are, ii)  how likely it is that this mode is misidentified in observations as an ordinary  AW and what the consequences of this could be, and, iii) if abundant, how important this mode  may be, in particular for heating in the solar plasma environment.

i) Examples of interpenetrating plasmas in space and astrophysics are numerous, like colliding astrophysical clouds, and  plasmas originating from the explosions of novae and supernovae  and moving through the surrounding
interstellar space.

 An obvious example of such  plasmas is
seen also in the solar atmosphere.  Observations\citep{bru, benz} describe  jets
generated in the transition region of the solar atmosphere
with the rate of about 24 events per second throughout the
solar atmosphere, and having a few thousand kilometers in
diameter. Their upward speed was around 400 km/s. Up-flows of plasma in
 spicules may reach any height in the
corona between $5\cdot 10^3$ km and $2\cdot 10^5$ km, with diameters
sometimes  greater than 1000 km.  They appear to be ubiquitous, $10^5$ events at any time and  covering a few percent
of the solar atmosphere. An opposite phenomenon, called
the coronal rain belongs to the same sort of phenomena \citep{benz}, where plasma performs
almost a free-fall speed between 50 and 100 km/s. So interpenetraing/permeating plasmas appear to be a rather
common phenomena in space plasmas.

Regarding the point ii), the real part of the obtained frequency
shows some peculiar features: for two critical values of the flow speed $v_0$, the frequency of P-solution (true forward mode) becomes the same as the AW frequency $\omega_s$ in the static plasma (the points denoted as $a$  and $b$ in Figs.~\ref{f1}, \ref{f2}). For yet another value of $v_0$ and for equal densities of the two plasmas, the wave frequency becomes equal to the AW frequency of the two plasmas combined $\omega_{s+f}$ (the point $d$ in Fig.~\ref{f1}). The second (N-solution) is with negative frequency only in the domain of small flow speed, while for greater  speed $v_0$  it is a forward mode as well. With such features it fits neither into the usual AW theory (where only one forward mode with the given polarization may exist), nor into the usual flow-driven mode $\omega= k v_0$, because the flow can only produce one (forward) mode, which is the P-solution discussed above. So the forward propagating N-solution is an intrinsic  result of coupling between  AW and FD modes, but it may have a positive frequency (including values close to $\omega_s$) and it can be misidentified as an ordinary AW whenever its frequency is positive. This implies that observers might deduce incorrect values for local plasma parameters by assuming it is an ordinary AW.

From practical side, rather  important is the situation described by the point $b$ because it describes a growing wave which  should easily develop and can consequently be observed.
On the other hand, from Eqs.~(\ref{e1}, \ref{e2}) it may be seen that the fluid motion in the presence of the flowing plasma (the term with $v_0$) is in the same direction as in the case of an ordinary AW. The transverse displacement is of the same magnitude as long as $v_0$ is of the same order as the phase speed. Thus,  both frequency and the fluid displacement  may be completely similar to the AW case (although clearly in this regime this is more the FD mode). The awareness of the existence  of this coupled AW-FD mode wave is thus of essential importance.

Regarding the issue iii), the AW and FD  modes are obviously closely interrelated. The resulting coupled mode has a clearly identified source, which is not always so for the standard AW. The features of both modes are similar, so quite generally,  eventual presence of this  electromagnetic flow-driven instability in the solar plasma environment can only strengthen the role of  the AW in the process of heating. According to the numbers on solar spicules\citep{dp2}, the speeds in spicules are up to 100 km/s, but the Alfv\'{e}n speed $v_{{\sss A}s}$ is always too big to have the instability. Even taking a rather weak field of 10 G and (too) high density $n_s=10^{17}$ m$^{-3}$  yields AW speed around 70 km/s, so the instability may work only in regions with weak field, and it could thus contribute to heating preferably in regions outside of strong magnetic structures where a proper model for heating is yet to be established.  However, the situation may be rather  different in case of the mentioned jets  \citep{bru} where the speed $v_0$ is far greater,  and the magnetic field may in principle be of the value given above or weaker. In view of  large spatial scales of these jets, the electromagnetic instability presented here may develop on quite large scales in horizontal direction, after a jet reaches  altitudes at which the conditions (\ref{e8}, \ref{e9}) become satisfied. So the total volumes within which the growing modes develop may be rather substantial.

The result presented in the work are relevant to numerous studies on the non-resonant stochastic heating of plasmas\citep{wang, wu, lu} by AW. In these studies this stochastic mechanism of heating has been discussed specifically in application to stellar atmospheres. The heating has various features that may explain observations, like better heating of heavier particles, better heating in the perpendicular direction with respect to the magnetic field, acceleration of ions in the parallel direction, and the obtained heating time\citep{lu} is of the order of $\pi/(k \vti)$.
 
 In application to the solar corona, for wavelengths of 1 and 100 km this gives heating times for protons roughly of the order of $1/200$ s and $1/2$ s,  respectively.  The obtained temperature increase\citep{lu} in parallel and perpendicular direction for protons was by factor 3.2 and 15, respectively,  and the achieved anisotropy was around $T_\bot/T_\|\simeq 5$. Protons are in average accelerated to the speed of around 0.15 Alfv\'{e}n speed. The anisotropy obtained for O$^{5+}$ ions  was 10-18. They estimated the heating time for O$^{5+}$ ions would be 10-40 minutes in the solar corona for waves  with 2 minutes wave-period.  The reported heating time\citep{wang}  was just a few gyro-rotation periods. It is  pointed out that a significant level of AW is needed for  theory to be realistic.  
 
 In the present work we have shown that plasma dynamics in interpenetrating plasmas remains similar to the ordinary AW case, while in the same time we provide a realistic energy source for such an AW heating. In other words, the mentioned condition about the required AW level is satisfied by the fact that we indeed have a growing mode. This all implies that the energy of the flowing plasma can directly be transformed into heat through this stochastic heating mechanism.

\end{document}